\documentclass[runningheads,a4paper]{llncs}
\usepackage{graphicx}
\usepackage{color}
\usepackage[numbers, sectionbib]{natbib}
\usepackage{hyperref}
\usepackage{amsmath,amssymb}
\usepackage{wasysym}
\usepackage{enumerate}
\usepackage[inline]{enumitem}
\urldef{\mailsa}\path|{k.mikhail, m.mazzara, l.safina, v.rivera}@innopolis.ru| 
\newcommand{\keywords}[1]{\par\addvspace\baselineskip
\noindent\keywordname\enspace\ignorespaces#1}

\begin{document}

\mainmatter

\title{Domain Objects and Microservices for \\Systems Development: a roadmap}
\author{Kizilov Mikhail\inst{1} \and Antonio Bucchiarone\inst{2} \and Manuel Mazzara\inst{1} \and \\ Larisa Safina\inst{1} \and Victor Rivera\inst{1}}

\institute{Innopolis University
\\
1, Universitetskaya Str., Innopolis, Russia, 420500\\
\mailsa\\
\url{https://www.university.innopolis.ru}
\and
Fondazione Bruno Kessler (FBK), Via Sommarive, 18, Trento, Italy,\\
bucchiarone@fbk.eu
}
\authorrunning{Kizilov Mikhail et al.}

\maketitle

\begin{abstract}
This paper discusses a roadmap to investigate \textit{Domain Objects} being an adequate formalism to capture the peculiarity of microservice architecture, and to support Software development since the early stages. It provides a survey of both \textit{Microservices} and \textit{Domain Objects}, and it discusses plans and reflections on how to investigate whether a modeling approach suited to adaptable service-based components can also be applied with success to the microservice scenario.

\keywords{Microservices, Domain Objects, Software Modeling}
\end{abstract}

\section{Introduction}\label{sec:intro}

The increasing complexity of modern software, which requires to be flexible and rapidly deployable, demands for new approaches to architectural design and system modeling. These approaches have to support developers from early stages and be able to produce quality software.

Innovative engineering is always looking for adequate instruments to model and verify software systems and support developers all along the development process in order to deploy correct software. Microservices~\cite{Dragoni2017} recently demonstrated to be an effective architectural paradigm to cope with scalability in a number of domain~\cite{DLLMMS2017}, including mission-critical systems~\cite{DDLM2017}. However, the paradigm still misses a conceptual model able to support engineers starting from the early phases of development. 

At the same time, Domain Objects (DO)~\cite{BucchiaroneSMPT15, BucchiaroneSMPT16} have been successfully used to model several case studies showing to be very effective in a service-based scenario and for composition of complex workflows of autonomous, heterogeneous and distributed services. Literature about service-workflow modeling is vast, in particular for B2B~\cite{YanMCU07}. However, Domain Objects are appearing in recent years as reference in the field. In this paper, we start an exploration of how development of microservice-based systems could be based on such approach. 

The paper is structured as follows. After this introduction, in Section~\ref{sec:micro} and Section~\ref{sec:DO} we will discuss the main concepts of Microservice and DO literature. In Section~\ref{sec:roadmap} we will discuss the main research question and the need for a diagrammatic notation, before finally presenting the roadmap.

\section{Microservices}\label{sec:micro}

Microservices~\cite{Dragoni2017} is an architectural style originating from Service-Oriented Architectures (SOAs)~\cite{mackenzie2006}. It was proposed to cope with the problems monolith applications have introduced, such as:

\begin{itemize}
\item complexity of monolith applications which complicates their maintainability;
\item impact of any part of the system changing have on the execution or redeployment of the whole system (any upgrade will call for system reboot);
\item limitations for system scalability (scaling the whole system instead of scaling only the parts experiencing the load);
\item constraints of using one technology or programming language.
\end{itemize}

The main idea is to structure systems by composing small independent building blocks communicating exclusively via message passing. These components are called \textit{microservices}, the term was first introduced at an architectural workshop in 2011 as a participants proposal of naming the new architectural pattern they have explored. Before the term was coined, microservices were called differently, e.g. Netflix named them ``Fine grained SOA'', showing that the microservices architecture is the nearest successor of SOA. However, microservices architecture can be distinct from SOA by some key characteristics, such as service size (relatively small with respect to services in SOA), service independence and bounded context. 

Each microservice is expected to implement a single \textit{business capability}, bringing benefits in terms of service maintainability and extendability. Since each microservice represents a single business capability, which is delivered and updated independently, discovering  bugs or adding a minor improvements do not have any impact on other services and on their releases.

One of the characteristic differentiating the new style from monolithic architectures and SOA is the emphasis on scalability. As microservices are implemented by independent instances, possible to be deployed on different hosts, natural distribution of the workload arises, making the system significantly more efficient and boosting the system availability. It can be also easily located which components of the system is
affected by high load which makes possible to scale them independently and with fine granularity without affecting the availability of other components. Microservices and their supporting environment (databases, libraries, etc.) can be packaged in containers  and deployed on any platform supporting the chosen container technology, they also can be easily replicated and dynamically scaled according to the current load. The ease of replication affects such quality as availability and robustness, since fault tolerance is ensured by using of possible redundant services. That all makes microservice architectures a good choice a system horizontally scaling is required. 

Microservices have seen their popularity blossoming with an explosion of concrete applications seen in real-life software~\cite{N15}. Several companies are involved in a major refactoring of their backend systems in order to improve scalability \cite{DLLMMS2017}. In~\cite{DDLM2017} a real world case study, concerning the migration of a mission critical system from an existing monolithic architecture to microservices, has been presented.

Such a notable success gave rise to academic and commercial interest, and ad-hoc programming languages arose to address the new architectural style \cite{MGZ14}. In principle, any general-purpose language could be used to program microservices. However, some of them are more oriented towards scalable applications and concurrency \cite{Guidi2017} . The Jolie programming language, for example, is based on the new paradigm and it allows to describe computation from a data-driven instead of process-driven perspective \cite{Safina2016}. As another advantage, Jolie has already a large community of users and developers \cite{Bandura16}

\section{Domain Objects}\label{sec:DO}
Internet of Services is the future of Internet focusing on real services rather than on software services. The main idea is to compose the available services on the Internet in value-added real services. The composition of such Service-based systems (SBSs) is not a trivial task, due to dynamic, context-aware, user-centric, and asset-based environments where they operate . Thus, new methodologies, techniques and tools are needed for this novel service composition \cite{BucchiaroneMPR17}. In addition, such SBSs should provide mechanisms and tools for the enactment, monitoring, adaptation, management of the delivered services \cite{Pistore2009}.

Design of such systems tends to have a lot of issues and requirements \cite{Marconi2012}. The SBS requires at the design time novel life-cycle that considers \emph{design for adaptation} as the first class concern of SBS and adds new iteration cycle at run-time to address adaptation needs on-the-fly. Also, to design such applications different alternatives to support service adaptation should be identified, such as \textit{adaptation mechanisms} and \textit{adaptation strategies} \cite{Bucchiarone2010}.

The main concept and design model of overall system based on \emph{Domain Objects}  has been presented in \cite{BucchiaroneSMPT16} and exploited in the development of various applications as Smart Mobility \cite{ICSOC2017}, Smart Logistics \cite{RaikBKMP12}, and Mobile Multi-robot Systems \cite{ICSOC2016}. The proposed approach based on the following main components of the system:

\textbf{Wrapping} component encapsulates the independent and heterogeneous services and present them as open, uniform and reliable services. In this context, a \textit{domain object} (DO) has been thought of as a uniform way to model autonomous and heterogeneous services at a level of abstraction that also allow for their easy interconnection through dynamic relations. Each DO has a partial view on the surrounding operational environment that is described by a set of concepts representing its \textit{domain knowledge}. 

The detailed structure of the DO has been presented in \cite{BucchiaroneSMPT15,BucchiaroneSMPT16}. The DO is modeled through two layers, namely, \textit{core layer} and \textit{fragments layer}. The core layer defines the \textit{internal behavior} of a DO. The fragments layer is represented through fragments \cite{Bucchiarone2012} which are exposed to the system and used by other DO to refine abstract activities (i.e., place holders) at run-time through incremental composition of different fragments. The incremental service composition realized by exploiting existing dynamic composition techniques such as presented in \cite{BucchiaroneMPR17}.

\textbf{Execution} Component takes in charge the  Orchestration and Choreography of services realized as DO. Orchestration represents control of the overall process flow, using  appropriate DOs and determine what steps to complete (i.e., abstract activities). In contrast, choreography used to compose higher-level services from existing orchestrated processes to track messages between these parties. In \cite{Peltz2003} these two concepts  illustrated by using two main standardized languages developed for web services orchestration and choreography, namely BPEL and WSDL. However, there are number of limitations associated with composition of services related to the assumption that designer knows the service to be composed during the design time. Moreover, such approach leads to strongly linked to particular service implementations. Therefore, proposed solutions are not adequate to dynamic service based environments.
In conclusion, an adaptive system is realized as a dynamic network of domain objects connected through a set of dependencies established through their runtime interactions by means of their offered/required functionalities. In such a system, each DO can self-adapt its behaviour according to the available services in the specific execution context and to the changes affecting its execution.

\textbf{Monitoring} 
An important feature of DO is the possibility of leaving the handling of extraordinary/improbable situations (e.g., context changes, availability of functionalities, improbable events) to run-time instead of analyzing all the extraordinary situations at design-time and embedding the corresponding recovery activities at execution time. 
These dynamic features rely on a shared \textit{domain model}, describing the operational environment of the system. The domain is defined through a set of \textit{domain properties}, each describing a particular aspect of the system domain (e.g., current location of a person, availability of a specific service or resource). A domain property may evolve as an effect of the execution of a fragment activity, which corresponds to the “normal” behavior of the domain (e.g., current location of a passenger is at the pickup point), but also as a result of exogenous changes (e.g., road blocked).

Process and fragments of a DO are modeled as \emph{Adaptable Pervasive Flows} (APFs) \cite{APFL}, an extension of traditional workflow languages (e.g., BPEL) which makes them suitable for adaptation and execution in dynamic pervasive environments.
In addition to classical workflow language constructs (e.g., input, output, data manipulation, complex control flow constructs), APFs allows to relate the process execution to the system domain by annotating activities with \textit{preconditions} and \textit{effects}. 
\textit{Preconditions} constrain the activity execution to specific domain configurations, and are used to catch violations in the expected behavior and trigger run-time adaptation. \textit{Effects} model the expected impact of the activity execution on the system domain, and are used to automatically reason on the consequences of fragment/process execution.

Activities can also be annotated with a \emph{compensation goal} that has to be fulfilled any time adaptation requires to roll-back the process instance and they have already been successfully executed.  
 
\textbf{Adaptation} component allows both for effectively dealing with domain changes and for reducing the degree of services coupling, since the interconnection among DOs is postponed from the design phase of the system to its execution. 
In order to resolve an adaptation need the framework offers a set of  \textit{adaptation mechanisms}: \emph{refinement}, \emph{local adaptation}, and \emph{compensation}. These mechanisms can be combined to form more complex mechanisms and strategies.

The \emph{refinement mechanism} is triggered whenever an abstract activity in a process instance needs to be refined. The aim of this mechanism is to automatically compose available fragments taking into account the goal associated to the abstract activity and the current domain configuration. 
The result of the refinement is an executable process that composes a set of fragments provided by other DOs in the system and, if executed, fulfills the goal of the abstract activity. 
Composed fragments may also contain abstract activities which requires further refinements during the process execution. This results in a multi-layer process execution model, where the top layer is the initial process of the entity and intermediate layers correspond to incremental refinements.

\textit{Local adaptation} aims at identifying a solution that allows re-starting the execution of a faulted process from a specific activity. To achieve this,  a composition of fragments is generated and its execution brings the system to a domain configuration satisfying the activity precondition.

The \emph{compensation mechanism} is used to dynamically compute a compensation process for a specific activity. The compensation process is a composition of fragments  whose execution fulfills the compensation goal.
The advantage of specifying activity compensation as a goal on the domain, rather than explicitly declaring the activities to be executed (e.g., as in BPEL), is in the possibility to dynamically compute the compensation process taking into account the specific execution domain. Secondly, the mechanism automatically generates different compensation processes depending on the status of the execution progress of the process. 

Different adaptation mechanisms can be combined to obtain more complex mechanisms. For instance,  \emph{re-refinement}  can be applied whenever a faulted activity belongs to the refinement of an abstract activity. The aim of this mechanism is to compensate all executed  activities of the refinement (through compensation mechanism) and to compute a new refinement (through refinement mechanism) that satisfies the goal of the abstract activity.

Another example of mechanisms composition is \emph{backward adaptation}. This mechanism aims at bringing the process instance back to a previous activity in the process that, given the new domain configuration, may allow for different execution decisions. This mechanism requires the compensation of all the activities that need to be rolled-back, and of bringing the system to a configuration where the precondition of the activity to be executed is satisfied (local adaptation).

\emph{Adaptation strategies} are defined by associating an ordered set of adaptation mechanisms to adaptation triggers. To give an example, a possible strategy could be the following: whenever an activity precondition is violated (adaptation trigger) search for a local adaptation, and, in case no solution is found, try backward adaptation within the same fragment composition, if this does not succeed, then apply the re-refinement mechanism. 

\subsection{Adaptive Service-based Systems with Domain Objects}
The SBS designed using DOs has capabilities to automatically adapt at runtime, through the monitoring of the execution environment and to solve the adaptation problems by combining fragments exposed in the system. The DO approach, offers a lightweight-model, with respect to the existing languages for service composition. It can be implemented in any object-oriented language (i.e., Java) and define both orchestration and choreography thanks to hierarchical organization of DO. For instance, several implementation in Java language have been presented: partial implementation of core concepts of DO \cite{BucchiaroneSMPT15} and Urban Mobility System Demonstrator - overall system with planning engine, where services modeled with DO \cite{DEMOCAS}. 

Unlike traditional systems where the behavior expected at run-time is specialized at design time, the approach based on DO allows the system to define dynamic behavior through partial definition of processes. This task is accomplished with abstract activities, which are refined when the context is known or discovered. The proposed design method for adaptive by design SBSs represents the effectiveness both to the wide range of changes that may occur in the system \cite{BucchiaroneSMPT15} and in terms of efficiency of refinement abstract activities and solve the AI planning problems \cite{BucchiaroneMPR17}.

\begin{figure}[tbh!]
\centering
     \includegraphics[width=1.0\columnwidth]{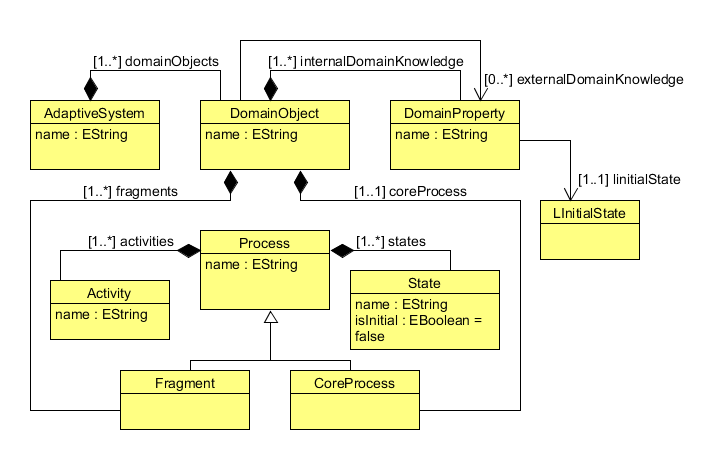}
  \caption{Excerpt of the Adaptive System metamodel.}
 \label{fig:adaptivesystemMM}
\end{figure}

An excerpt of the Adaptive System model \footnote{For the sake of space, the metamodels are not presented completely. The reader is referred to https://github.com/das-fbk/CAS-DSL for the complete metamodels.}, based on DO is shown in Figure~\ref{fig:adaptivesystemMM}.
An \texttt{AdaptiveSystem} is a composition of \texttt{DomainObject}s, each of which including a \texttt{CoreProcess}, \texttt{Fragment}s, and \texttt{DomainProperty}s. It is worth noting that the multiplicity boundaries put constraints on the well-formedness of an Adaptive System model. Notably, there must be at least a \texttt{DomainObject}, and each \texttt{DomainObject} must contain one unique \texttt{CoreProcess}. The relationships between domain objects and domain properties establish that a domain property represents \texttt{internaldomainknowledge} if defined within the \texttt{DomainObject} (composition relation), whereas it represents \texttt{externaldomainknowledge} if referred to by a simple association. 

Both processes and domain properties can be reduced to state transition systems \cite{BucchiaroneSMPT15}. From a modelling point-of-view, the only difference between the two is that for processes (both core and fragments) there is no notion of initial state, or better, it is possible to set multiple states as initial through a Boolean attribute (see \texttt{isInitial} in \texttt{State}). On the contrary, a \texttt{DomainProperty} must have a \texttt{LInitialState}, as constrained by the multiplicity boundaries of the \texttt{linitialstate} relationship.

\section{Research Objectives}\label{sec:roadmap}

In this section we will describe the key research objectives towards the utilization of DO as model of the microservice architecture and we will identify step towards reaching the objectives.

\subsection{Research question}

The DO approach demonstrated to be suitable to describe adaptable service-based components.  How can we extend its applicability to Microservices? The research question can be formulated as follow : \textit{is the DO formalism suited to describe software system to be built according to the microservice architecture?} In other words, is the formalism over-expressive or under-expressive? Are the features of the microservice architecture well represented by the formalism or, to the contrary, the modeling tool is overcomplex for a relatively simple architectural style?

In general, how is it possible to answer this research question, and how is it possible to provide sufficient evidence in order to support any claim in this area? Our strategy is evidence-based via a case study. The roadmap includes the choice of an applicative scenario on which to experiment with modeling. We identified this scenario as coming from Internet of Things (IoT), in particular the one described in \cite{Salikhov2016a,Salikhov2016b}.

While some of the scenario previously modeled by DO may be described as over-complex, the one we have chosen is simple and it has been implemented by ourself in the university building. This provides a control over implementation and deployment, and an immediate feedback between modeling and development. In fact, we can adapt the implementation as needed in the same way we can adapt the model, in order to see how they can fit each other. The realization of the case study, both in terms of modeling and deployment will represent an opportunity to discuss and answer to the aforementioned research question.

\subsection{Diagrammatic notation}

A second objective in our roadmap is the development of a diagrammatic representation of DO which is consistent with the mathematical formulation. This diagrammatic representation will be experimented again via the case study that can test its suitability to the microservice architecture. As demonstrated by the long experience of UML and ER diagrams, for example, valid theory and mathematical modeling tools have reached widespread adoption when coupled with visual tools (and software able to support creation and drawing). Visual tools are fundamental in the requirements engineering phase and in the interaction with customers, allowing early mutual understanding of the system under construction.

\begin{figure}[h]
\centering
     \includegraphics[width=0.7\textwidth]{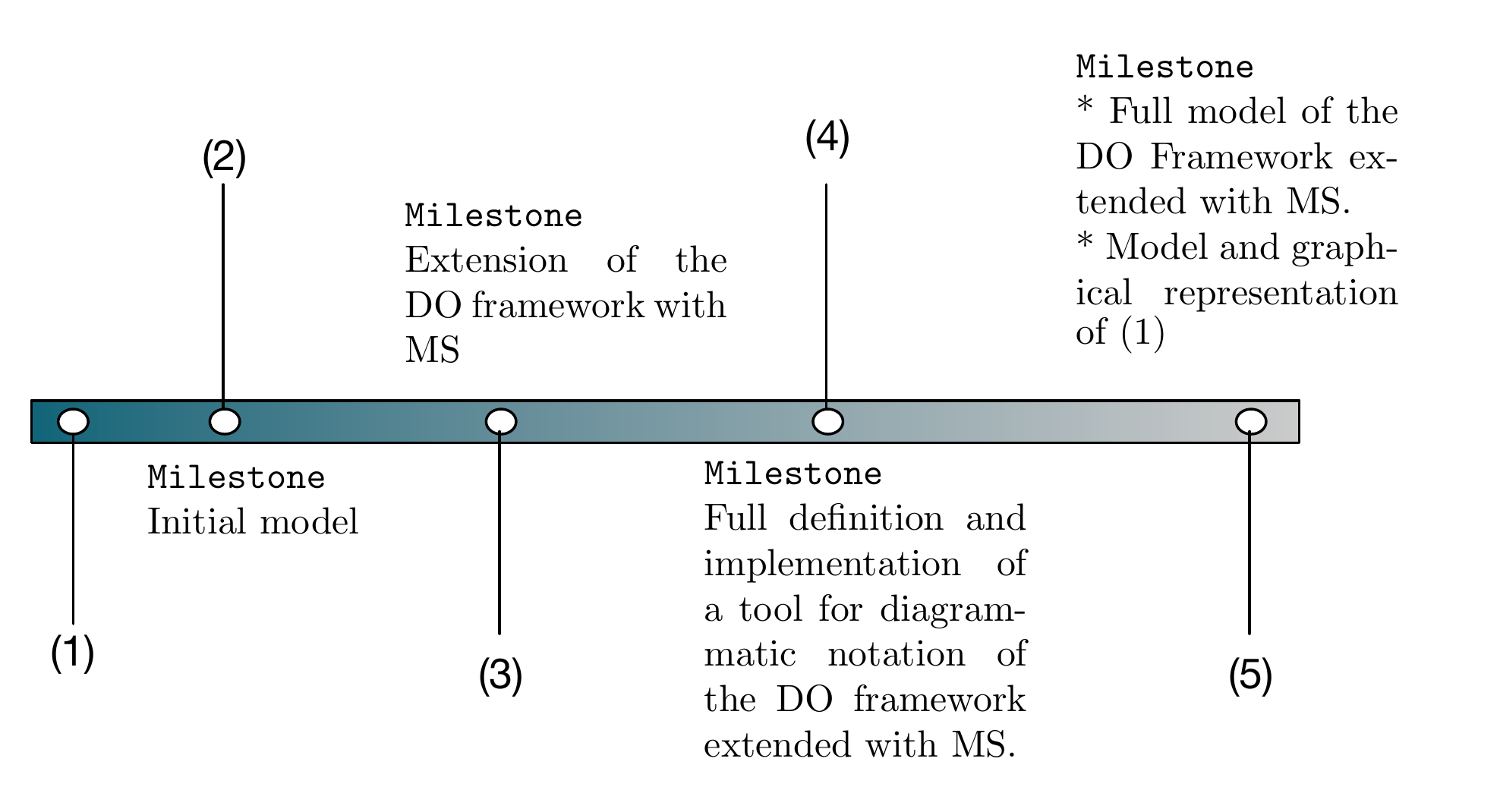}
  \caption{Roadmap plan and milestones.}
 \label{fig:milestone}
\end{figure}

\subsection{Roadmap}

Our future work is planned as follows (Figure \ref{fig:milestone} depicts our plan with specific milestones):


\begin{enumerate}
\item Identify a case study in the IoT area on which we have control over deployment;
\item Analyzing the case study and experimenting with modeling;
\item Answering the aforementioned research question, therefore identifying how to extend the DO framework to be used with Microservices;
\item During modeling identify needs for a diagrammatic notation and fine tune it;
\item Coming out with a full modeling of the case study and its corresponding visual representation.
\end{enumerate}

\emph{Docker} is a popular technology these days \cite{GDM2017}. This success makes it impossible to investigate microservice architecture and tools to model it without taking this technology into account. A further development of the research will have to investigate a mapping between DO and Docker containers. In the future it will also be necessary to test the suitability of the diagrammatic notation and to develop tools in order to support software architects in the the modeling.


\bibliographystyle{ieeetr}
\bibliography{references}

\begin{thebibliography}{10}

\bibitem{Dragoni2017}
N.~Dragoni, S.~Giallorenzo, A.~Lluch-Lafuente, M.~Mazzara, F.~Montesi,
  R.~Mustafin, and L.~Safina, ``Microservices: yesterday, today, and
  tomorrow,'' in {\em Present and Ulterior Software Engineering}, Springer,
  2017.

\bibitem{DLLMMS2017}
N.~Dragoni, I.~Lanese, S.~T. Larsen, M.~Mazzara, R.~Mustafin, and L.~Safina,
  ``Microservices: How to make your application scale,'' in {\em A.P. Ershov
  Informatics Conference (the PSI Conference Series, 11th edition)}, Springer,
  2017.

\bibitem{DDLM2017}
N.~Dragoni, S.~Dustdar, S.~T. Larsen, and M.~Mazzara, ``Microservices:
  Migration of a mission critical system,''
\newblock \url{https://arxiv.org/abs/1704.04173}.

\bibitem{BucchiaroneSMPT15}
A.~Bucchiarone, M.~D. Sanctis, A.~Marconi, M.~Pistore, and P.~Traverso,
  ``Design for adaptation of distributed service-based systems,'' in {\em
  Service-Oriented Computing - 13th International Conference, {ICSOC} 2015,
  Goa, India, November 16-19, 2015, Proceedings}, pp.~383--393, 2015.

\bibitem{BucchiaroneSMPT16}
A.~Bucchiarone, M.~D. Sanctis, A.~Marconi, M.~Pistore, and P.~Traverso,
  ``Incremental composition for adaptive by-design service based systems,'' in
  {\em {IEEE} International Conference on Web Services, {ICWS} 2016, San
  Francisco, CA, USA, June 27 - July 2, 2016}, pp.~236--243, 2016.

\bibitem{YanMCU07}
Z.~Yan, M.~Mazzara, E.~Cimpian, and A.~Urbanec, ``Business process modeling:
  Classifications and perspectives,'' in {\em Business Process and Services
  Computing: 1st International Working Conference on Business Process and
  Services Computing, {BPSC} 2007, September 25-26, 2007, Leipzig, Germany.},
  p.~222, 2007.

\bibitem{mackenzie2006}
M.~MacKenzie {\em et~al.}, ``Reference model for service oriented architecture
  1.0,'' {\em OASIS Standard}, vol.~12, 2006.

\bibitem{N15}
S.~Newman, {\em Building microservices}.
\newblock O'Reilly Media, Inc., 2015.

\bibitem{MGZ14}
F.~Montesi, C.~Guidi, and G.~Zavattaro, ``{Service-Oriented Programming with
  Jolie},'' in {\em Web Services Foundations}, pp.~81--107, Springer, 2014.

\bibitem{Guidi2017}
C.~Guidi, I.~Lanese, M.~Mazzara, and F.~Montesi, ``Microservices: a
  language-based approach,'' in {\em Present and Ulterior Software
  Engineering}, Springer, 2017.

\bibitem{Safina2016}
L.~Safina, M.~Mazzara, F.~Montesi, and V.~Rivera, ``Data-driven workflows for
  microservices (genericity in jolie),'' in {\em Proc. of The 30th IEEE
  International Conference on Advanced Information Networking and Applications
  (AINA), 2016}.

\bibitem{Bandura16}
A.~Bandura, N.~Kurilenko, M.~Mazzara, V.~Rivera, L.~Safina, and A.~Tchitchigin,
  ``Jolie community on the rise,'' in {\em SOCA}, pp.~40--43, {IEEE} Computer
  Society, 2016.

\bibitem{BucchiaroneMPR17}
A.~Bucchiarone, A.~Marconi, M.~Pistore, and H.~Raik, ``A context-aware
  framework for dynamic composition of process fragments in the internet of
  services,'' {\em J. Internet Services and Applications}, vol.~8, no.~1,
  pp.~6:1--6:23, 2017.

\bibitem{Pistore2009}
M.~Pistore, P.~Traverso, M.~Paolucci, and M.~Wagner, ``{From Software Services
  to a Future Internet of Services},'' pp.~183--192, 2009.

\bibitem{Marconi2012}
A.~Marconi, A.~Bucchiarone, K.~Bratanis, A.~Brogi, J.~C\'{a}mara, D.~Dranidis,
  H.~Giese, R.~Kazhamiakin, R.~de~Lemos, C.~C. Marquezan, and A.~Metzger,
  ``Research challenges on multi-layer and mixed-initiative monitoring and
  adaptation for service-based systems,'' in {\em Proceedings of the First
  International Workshop on European Software Services and Systems Research:
  Results and Challenges}, S-Cube '12, (Piscataway, NJ, USA), pp.~40--46, IEEE
  Press, 2012.

\bibitem{Bucchiarone2010}
A.~Bucchiarone, C.~Cappiello, E.~Di~Nitto, R.~Kazhamiakin, V.~Mazza, and
  M.~Pistore, {\em Design for Adaptation of Service-Based Applications: Main
  Issues and Requirements}, pp.~467--476.
\newblock Berlin, Heidelberg: Springer Berlin Heidelberg, 2010.

\bibitem{ICSOC2017}
A.~Bucchiarone, M.~D. Sanctis, and A.~Marconi, ``{ATLAS}: A world-wide travel
  assistant exploiting service-based adaptive technologies,'' in {\em
  Service-Oriented Computing - 15th International Conference, {ICSOC} 2017,
  Málaga, Spain, November 13-16, 2017, To Appear}, 2017.

\bibitem{RaikBKMP12}
H.~Raik, A.~Bucchiarone, N.~Khurshid, A.~Marconi, and M.~Pistore,
  ``Astro-captevo: Dynamic context-aware adaptation for service-based
  systems,'' in {\em Eighth {IEEE} World Congress on Services, {SERVICES} 2012,
  Honolulu, HI, USA, June 24-29, 2012}, pp.~385--392, 2012.

\bibitem{ICSOC2016}
A.~Bucchiarone, M.~D. Sanctis, and A.~Marconi, ``Decentralized dynamic
  adaptation for service-based collective adaptive systems,'' in {\em
  Service-Oriented Computing - 15th International Conference, ASOCA Workshop at
  ICSOC 2016, October 10-13, Banff, Alberta, Canada, (To Appear)}, 2016.

\bibitem{Bucchiarone2012}
A.~Bucchiarone, A.~Marconi, M.~Pistore, and H.~Raik, ``Dynamic adaptation of
  fragment-based and context-aware business processes,'' {\em Proceedings -
  2012 IEEE 19th International Conference on Web Services, ICWS 2012},
  no.~June, pp.~33--41, 2012.

\bibitem{Peltz2003}
C.~Peltz, ``{Web Services Orchestration and Composition},'' {\em Computer},
  vol.~36, no.~10, pp.~46--52, 2003.

\bibitem{APFL}
A.~Bucchiarone, A.~Lluch{-}Lafuente, A.~Marconi, and M.~Pistore, ``A
  formalisation of adaptable pervasive flows,'' in {\em Web Services and Formal
  Methods, 6th International Workshop, {WS-FM} 2009, Bologna, Italy, September
  4-5, 2009, Revised Selected Papers}, pp.~61--75, 2009.

\bibitem{DEMOCAS}
A.~Bucchiarone, M.~D. Sanctis, A.~Marconi, and A.~Martinelli, ``{DeMOCAS:}
  domain objects for service-based collective adaptive systems,'' in {\em
  Service-Oriented Computing - 15th International Conference, Demo paper at
  ICSOC 2016, October 10-13, Banff, Alberta, Canada, (To Appear)}, 2016.

\bibitem{Salikhov2016a}
D.~Salikhov, K.~Khanda, K.~Gusmanov, M.~Mazzara, and N.~Mavridis,
  ``Microservice-based iot for smart buildings,'' in {\em WAINA}, 2017.

\bibitem{Salikhov2016b}
D.~Salikhov, K.~Khanda, K.~Gusmanov, M.~Mazzara, and N.~Mavridis, ``Jolie good
  buildings: Internet of things for smart building infrastructure supporting
  concurrent apps utilizing distributed microservices,'' in {\em CCIT},
  pp.~48--53, 2016.

\bibitem{GDM2017}
A.~Giaretta, N.~Dragoni, and M.~Mazzara, ``Joining jolie to docker -
  orchestration of microservices on a containers-as-a-service layer,''
\newblock \url{https://arxiv.org/abs/1709.05635}.

\end{thebibliography}

\end{document}